
\input harvmac
\def\psheader#1{}


\font\blackboard=msbm10 \font\blackboards=msbm7
\font\blackboardss=msbm5
\newfam\black
\textfont\black=\blackboard
\scriptfont\black=\blackboards
\scriptscriptfont\black=\blackboardss
\def\blackb#1{{\fam\black\relax#1}}



%
\def\BC{{\blackb C}} 
\def\BR{{\blackb R}} 
\def\BZ{{\blackb Z}} 
\def\BP{{\blackb P}}


%
\font\mathbold=cmmib10 \font\mathbolds=cmmib7
\font\mathboldss=cmmib5
\newfam\mbold
\textfont\mbold=\mathbold
\scriptfont\mbold=\mathbolds
\scriptscriptfont\mbold=\mathboldss
\def\bi{\fam\mbold\relax}


%
\font\gothic=eufm10 \font\gothics=eufm7
\font\gothicss=eufm5
\newfam\gothi
\textfont\gothi=\gothic
\scriptfont\gothi=\gothics
\scriptscriptfont\gothi=\gothicss

%

\def\tfig#1{Fig.~\the\figno\xdef#1{Fig.~\the\figno}\global\advance\figno by1}
\def\figI{I}
%
\newdimen\tempszb \newdimen\tempszc \newdimen\tempszd \newdimen\tempsze
\ifx\figflag\figI
\input epsf
%
\def\epsfsize#1#2{\expandafter\epsfxsize{
 \tempszb=#1 \tempszd=#2 \tempsze=\epsfxsize
     \tempszc=\tempszb \divide\tempszc\tempszd
     \tempsze=\epsfysize \multiply\tempsze\tempszc
     \multiply\tempszc\tempszd \advance\tempszb-\tempszc
     \tempszc=\epsfysize
     \loop \advance\tempszb\tempszb \divide\tempszc 2
     \ifnum\tempszc>0
        \ifnum\tempszb<\tempszd\else
           \advance\tempszb-\tempszd \advance\tempsze\tempszc \fi
     \repeat
\ifnum\tempsze>\hsize\global\epsfxsize=\hsize\global\epsfysize=0pt\else\fi}}
\epsfverbosetrue
\psheader{fig3.pro}       
\fi
%

%
%
%
%

\def\ifigure#1#2#3#4{
\midinsert
\vbox to #4truein{\ifx\figflag\figI
\vfil\centerline{\epsfysize=#4truein\epsfbox{#3}}\fi}
\baselineskip=12pt
\narrower\narrower\noindent{\bf #1:} #2
\endinsert
}
%
%
\def\ifigures#1#2#3#4#5#6#7#8{
\midinsert
\centerline{
\hbox{\vbox{
\divide\hsize by 2
\vbox to #4truein{\ifx\figflag\figI
\vfil\centerline{\epsfysize=#4truein\epsfbox{#3}}\fi}
\baselineskip=12pt
\narrower\narrower\noindent{\bf #1:} #2
}}\qquad
\hbox{\vbox{
\divide\hsize by 2
\vbox to #8truein{\ifx\figflag\figI
\vfil\centerline{\epsfysize=#8truein\epsfbox{#7}}\fi}
\baselineskip=12pt
\narrower\narrower\noindent{\bf #5:} #6
}}}
\endinsert
}


\def\appendix#1#2{\global\meqno=1\global\subsecno=0\xdef\secsym{\hbox{#1:}}
\bigbreak\bigskip\noindent{\bf Appendix #1: #2}\message{(#1: #2)}
\writetoca{Appendix {#1:} {#2}}\par\nobreak\medskip\nobreak}

%

\def\fourpt{\hbox{{$\rangle \kern-.25em \langle$}}} 
\def\tree{\hbox{{$\rangle \kern-.5em - \kern-.5em \langle$}}}
 %

\def\CM{{\cal M}} \def\CW{{\cal W}}

\def\ET{{\rm End(T)}}
\def\ex#1{{\rm e}^{#1}}                 

\def\cp#1{{\BC{\rm P}^{#1}}}

\def\sdp{{\blackb n}}

\def\Ka{K\"ahler}

\def\LG{Landau-Ginzburg}

\def\ql{{\bi q}}
\def\qr{\overline{\bi q}}

\noblackbox

\Title{\vbox{\hbox{PUPT--1464}\hbox{\tt hep-th@xxx/9406091}}}
{\vbox{\centerline{Quantum Symmetries and }
\medskip
\centerline{Stringy Instantons $^\star$}
}}

\centerline{Jacques Distler and Shamit Kachru$^\dagger$}\smallskip
\centerline{Joseph Henry Laboratories}
\centerline{Princeton University}
\centerline{Princeton, NJ \ 08544 \ USA}
\bigskip\bigskip

\footnote{}{{\parindent=-5pt\par $\dagger$Address
after June 30: Department of Physics, Harvard University, Cambridge, MA
02138.}}
\footnote{}{{\parindent=-5pt\par $\star$
\vtop{
\hbox{Email: {\tt distler@puhep1.princeton.edu}, {\tt
kachru@puhep1.princeton.edu} .}
\hbox{Research supported by NSF grant PHY90-21984, and the
A.~P.~Sloan Foundation.}
     }     }}

The quantum symmetry of many \LG\ orbifolds appears to be
broken by Yang-Mills instantons.
However,
isolated Yang-Mills instantons are not solutions of string theory: They
must be accompanied by
gauge anti-instantons, gravitational instantons, or topologically
non-trivial configurations of the $H$ field.
We check that the configurations permitted in string theory do
in fact preserve the quantum symmetry, as a result of non-trivial
cancellations between symmetry breaking effects due to the various types
of instantons.   These cancellations indicate
new constraints on \LG\ orbifold spectra
and require that the dilaton modulus
mix with the twisted moduli in some \LG\ compactifications.
We point out that one can find similar constraints
at all fixed points
of the modular group of the moduli space of vacua.

\Date{June 1994}                 

\lref\Rohm{R. Rohm and E. Witten, ``The Antisymmetric Tensor Field in
Superstring Theory,''~{\it Ann. Phys.}~{\bf 170} (1986) 454.}
\lref\Schwarz{J. Schwarz, ``Does String Theory have a Duality Symmetry
Relating Strong and Weak Coupling?,'' talk presented at Strings 93,
Berkeley, {\tt hep-th/9307121}.}
\lref\Duff{S.M. Christensen and M.J. Duff, ``New Gravitational Index
Theorems and Super Theorems,'' {\it Nucl. Phys.} {\bf B154} (1979)
301.}
\lref\Asen{A. Sen, ``Strong-Weak Coupling Duality in Four-Dimensional
String Theory,'' TIFR preprint, {\tt hep-th/9402002}.}
\lref\Hanson{See for example
T. Eguchi, P. Gilkey, and A. Hanson, ``Gravitation, Gauge
Theories and Differential Geometry,'' {\it Phys. Rep.} {\bf 66} (1980)
213-393, especially p.333.}

\lref\Sei{N. Seiberg, ``Exact Results on the Space of Vacua of
Four-Dimensional SUSY Gauge Theories,'' Rutgers preprint, {\tt
hep-th/9402044}.}
\lref\Gepner{D. Gepner, ``String Theory on Calabi-Yau Manifolds: The
Three Generation Case,'' Princeton preprint, December 1987.}
\lref\Schimmrigk{R. Schimmrigk, ``A New Construction of a
Three-Generation Calabi-Yau Manifold,'' {\it Phys. Lett.} {\bf 193B}
(1987) 175.}
\lref\Killing{T.P. Killingback, ``World-Sheet Anomalies and Loop
Geometry,'' {\it Nucl. Phys.} {\bf B288} (1987) 578.}
\lref\WitGlob{E. Witten, ``Global Anomalies in String Theory,''
in Proceedings of Argonne Symposium on Anomalies, Geometry, and Topology
(1985).}
\lref\DKtwo{J. Distler and S. Kachru, ``Singlet Couplings and (0,2)
Models,'' Princeton preprint PUPT-1465 (1994).}

\lref\Gt{G. 't Hooft, ``Symmetry Breaking through Bell-Jackiw Anomalies,''
{\it Phys. Rev. Lett.} {\bf 37} (1976) 8 \semi
``Computation of the Quantum Effects due to a Four-Dimensional
Pseudoparticle,'' {\it Phys. Rev.} {\bf D14} 1976) 3432.}
\lref\Freed{D.S. Freed, ``Determinants, Torsion, and Strings,''
{\it Comm. Math. Phys.} {\bf 107} (1986) 483.}
\lref\Callan{
A. Strominger, ``Heterotic Solitons,'' {\it Nucl. Phys.} {\bf B343}
(1990) 167 \semi
C. Callan, J. Harvey, and A. Strominger, ``Supersymmetric String
Solitons,'' lectures delivered at the 1991 Trieste Spring School, {\tt
hep-th/9112030}.}
\lref\BanksD{T. Banks and M. Dine, ``Note on Discrete Gauge Anomalies,''
{\it Phys. Rev.} {\bf D45} (1992) 1424, {\tt hep-th/9109045.}}

\lref\Leigh{M. Dine, R.G. Leigh, and D.A. MacIntire, ``Discrete
Gauge Anomalies in String Theory,'' Santa Cruz preprint, {\tt
hep-th/9307152}. }
\lref\WitMin{``On the Landau-Ginzburg Description of N=2 Minimal
Models,'' IAS preprint, {\tt hep-th/9304026}.}

\lref\AspSmall{P. Aspinwall, B. Greene, and D. Morrison, ``Measuring
Small Distances in N=2 Sigma Models,'' IAS preprint,
{\tt hep-th/9311042}.}
\lref\WitSome{E. Witten, ``Some Properties of O(32) Superstrings,''
{\it Phys. Lett.} {\bf 149B} (1984) 351.}
\lref\rdualanom{J. Derendinger, S. Ferrara, C. Kounnas and F. Zwirner,
Phys. Lett. {\bf 271B} (1991) 307\semi
J. Louis, SLAC-PUB-5527 (1991)\semi
G. Lopes Cardoso and B. Ovrut, Nucl. Phys. {\bf B369} (1992) 351\semi
L. Ib\'a\~nez and D. L\"ust, Phys. Lett. {\bf 267B} (1991) 51\semi
L. Ib\'a\~nez and D. L\"ust, {\it Nucl.  Phys.} {\bf B382} (1992) 305.}

\lref\SpecialGeo{B. de Wit, P. Lauwers and A. van Pr\oe yen, Nucl. Phys. {\bf
B255} (1985) 569.}
\lref\ranspec{S. Cecotti, S. Ferrara and L. Girardello, Int. J. Mod. Phys.
{\bf A4} (1989) 2475\semi
L. Castellani, R. D'Auria and S. Ferrara, Class. Quantum Grav. {\bf 7} (1990)
1767.}
\lref\CandMod{P. Candelas and X. De la Ossa, ``Moduli
Space of Calabi-Yau Manifolds,'' {\it Nucl. Phys.} {\bf B355} (1991)
455.}
\lref\OldEd{E. Witten, ``Symmetry Breaking Patterns in Superstring
Models,'' {\it Nucl. Phys.} {\bf B258} (1985) 75.}
\lref\DKL{L. Dixon, V. Kaplunovsky and J. Louis, Nucl. Phys. {\bf B329}
(1990) 27.}
 \lref\StromSpec{A. Strominger, ``Special Geometry,''
 {\it Comm. Math. Phys.} {\bf 133}
(1990) 163.}
\lref\PerStrom{V. Periwal and A. Strominger, Phys. Lett. {\bf B335} (1990)
261.}
\lref\TopAntiTop{S. Cecotti and C. Vafa, Nucl. Phys. {\bf B367} (1991) 359.}
\lref\AandB{E. Witten, in ``Proceedings of the Conference on Mirror Symmetry",
MSRI (1991).}
\lref\LVW{W. Lerche, C. Vafa and N. Warner, Nucl. Phys. {\bf B324} (1989)
427.} 
\lref\twisted{E. Witten. Comm. Math. Phys. {\bf 118} (1988) 411\semi
E. Witten, Nucl. Phys. {\bf B340} (1990) 281\semi
T. Eguchi and S.-K. Yang , Mod. Phys. Lett. {\bf A5} (1990) 1693.}
 \lref\GVW{B. Greene, C. Vafa and N. Warner,
``Calabi-Yau Manifolds and Renormalization Group Flows,''
 {\it Nucl. Phys.} {\bf B324} (1989)
371.}
\lref\Grisaru{M. Grisaru, A. Van de Ven and D. Zanon, Phys. Lett. {\bf 173B}
(1986) 423.}
\lref\DSWW {M. Dine, N. Seiberg, X.G. Wen and E. Witten,
``Non-Perturbative Effects on the String World Sheet I,'' {\it Nucl.
Phys.}~{\bf B278} (1986) 769, ``Non-Perturbative Effects on the String
World Sheet II,'' {\it Nucl. Phys.}~{\bf B289} (1987) 319. }
\lref\DixonRev{L. Dixon, in ``Proceedings of the 1987 ICTP Summer Workshop in
High Energy Physics and Cosmology", ed G. Furlan, {\it et. al.}}
\lref\Exact{J. Distler and B. Greene,
``Some Exact Results on the Superpotential from Calabi-Yau
Compactifications,''
{\it Nucl. Phys.}
 {\bf B309} (1988) 295.}
\lref\twozero{J. Distler and B. Greene,
``Aspects of (2,0) String Compactifications,''
{\it Nucl. Phys.} {\bf B304} (1988) 1.}
\lref\AspMor{P. Aspinwall and D. Morrison, ``Topological Field Theory and
Rational Curves," Comm. Math. Phys. {\bf 151} (1993) 245.}
\lref\CandMir{P. Candelas, X. de la Ossa, P. Green and L. Parkes,
``A Pair of Calabi-Yau Manifolds as an Exactly Soluble Superconformal
Theory,''
{\it Nucl.
Phys.} {\bf B359} (1991) 21.}
\lref\Kutconf{D. Kutasov, ``Geometry on the space of conformal field
theories and contact terms,'' Phys. Lett. {\bf B220} (1989) 153.}%
\lref\GrSei{M. Green and N. Seiberg, ``Contact interactions in
superstring theory," Nucl. Phys. {\bf B299} (1988) 559.}%
\lref\WilZee{F. Wilczek and A. Zee, Phys. Rev. Lett. {\bf 52} (1984) 2111.}
\lref\Banks{T. Banks, L. Dixon,
D.Friedan and E. Martinec, ``Phenomenology and Conformal Field Theory or
Can String Theory Predict the Weak Mixing Angle?,'' {\it Nucl. Phys.}
{\bf B299} (1988) 613.}
\lref\Hodge{P. Griffiths, Periods of Integrals on Algebraic manifolds I,II,
Am. J. Math. {\bf 90} (1970) 568,805\semi
R. Bryant and P. Griffiths, in ``Progress in Mathematics {\bf 36}"
(Birkh\"auser, 1983) 77.}
\lref\Odake{S. Odake, Mod. Phys. Lett. {\bf A4} (1989) 557\semi
S. Odake, Int. Jour. Mod. Phys. {\bf A5} (1990) 897\semi
T. Eguchi, H. Ooguri, A. Taormina and S-K. Yang, Nucl. Phys. {\bf B315}
(1989) 193.}
\lref\GS{M.B. Green and J.H. Schwarz, ``Anomaly Cancellations in
Supersymmetric D=10 Gauge Theory Require SO(32),'' {\it Phys. Lett.}
{\bf 149B}
(1984) 117.}

\lref\GSW{M. Green, J. Schwarz and E. Witten, ``Superstring theory, vol. II"
(Cambridge University Press, 1987).}
\lref\CHSW{P. Candelas, G. Horowitz, A. Strominger and E. Witten,
``Vacuum Configurations for Superstrings,'' {\it Nucl. Phys.} {\bf
B258} (1985) 46.}
\lref\NR{L. Alvarez-Gaum\'e, S. Coleman and P. Ginsparg, Comm. Math. Phys.
{\bf 103} (1986) 423.}
\lref\manin{D. Leites, ``Introduction to the theory of supermanifolds", Russ.
Math. Surveys {\bf 35} (1980) 3\semi
Yu. Manin, ``Gauge field theory and complex geometry", (Springer, 1988).}
\lref\Reviews{
J. Schwarz, ``Superconformal symmetry in string theory", lectures
at the 1988 Banff Summer Institute on Particles and Fields (1988)\semi
D. Gepner, ``Lectures on N=2 string theory",
lectures at the 1989 Trieste Spring School (1989)\semi
N. Warner, ``Lectures on N=2 superconformal theories and singularity theory",
lectures at the 1989 Trieste Spring School (1989)\semi
B. Greene, ``Lectures on string theory in four dimensions",
lectures at the 1990 Trieste Spring School (1990)\semi
S. Yau (editor), ``Essays in Mirror Manifolds",  Proceedings of the Conference
on Mirror Symmetry, MSRI (International Press, 1992).
}
\lref\Trieste{J. Distler, ``Notes on N=2 $\sigma$-models," lectures at the
1992 Trieste Spring School (1992), {\tt hep-th/9212062.}}
\lref\Seiberg{N. Seiberg, Nucl. Phys. {\bf B303} (1988) 286.}
\lref\Dubrovin{B. Dubrovin, ``Geometry and integrability of
topological--antitopological fusion", INFN preprint, INFN-8-92-DSF (1992).}
\lref\GPM{B. Greene, D. Morrison, and R. Plesser, in preparation.}
\lref\GPmirror{B. Greene and R. Plesser, Nucl. Phys. {\bf B338} (1990) 15.}
\lref\phases{E. Witten, ``Phases of N=2 Theories in Two Dimensions,"
{\it Nucl. Phys.} {\bf B403} (1993) 159, {\tt hep-th/9301042}.}
\lref\Vafa{C. Vafa, ``String Vacua and Orbifoldized LG models,''
{\it Mod. Phys. Lett.} {\bf A4} (1989) 1169.}
\lref\Ken{K. Intriligator and C. Vafa, ``Landau-Ginzburg Orbifolds,''
{\it Nucl. Phys.} {\bf B339} (1990) 95.}
\lref\Us{S. Kachru and E. Witten, ``Computing The Complete Massless
Spectrum Of A Landau-Ginzburg Orbifold,''
{\it Nucl. Phys.} {\bf B407} (1993) 637, {\tt hep-th/9307038}.}
\lref\WitMin{E. Witten, ``On the Landau-Ginzburg Description of N=2
Minimal Models,'' IAS preprint
IASSNS-HEP-93/10, {\tt hep-th/9304026}.}
\lref\Fre{P. Fr\'e, F. Gliozzi, M. Monteiro, and A. Piras, ``A
Moduli-Dependent Lagrangian For (2,2) Theories On Calabi-Yau n-Folds,''
{\it Class. Quant. Grav.} {\bf 8} (1991) 1455; P. Fr\'e, L. Girardello,
A. Lerda, and P. Soriani, ``Topological First-Order Systems With
Landau-Ginzburg Interactions,'' {\it Nucl. Phys.} {\bf B387} (1992)
333, {\tt hep-th/9204041}.}
\lref\Greene{B.R. Greene, ``Superconformal Compactifications in Weighted
Projective Space,'' {\it Comm. Math. Phys.} {\bf 130} (1990) 335.}
\lref\fendley{P. Fendley and K. Intriligator, ``Central Charges
Without Finite Size Effects,'' Rutgers preprint RU-93-26, {\tt
hep-th/9307101}.}
\lref\OldWit{E. Witten, ``New Issues in Manifolds of SU(3) Holonomy,''
 {\it Nucl. Phys.} {\bf B268} (1986) 79.}
\lref\Pasquinu{S. Cecotti, L. Girardello, and A. Pasquinucci,
``Non-perturbative Aspects and Exact Results for the N=2 Landau-Ginzburg
Models,'' {\it Nucl. Phys.}~{\bf B338} (1989) 701, ``Singularity
Theory and N=2 Supersymmetry,'' {\it Int. J. Mod. Phys.} {\bf A6} (1991)
2427.}
\lref\KT{A. Klemm and S. Theisen, ``Mirror Maps and Instanton Sums for
Complete Intersections in Weighted Projective Space,'' Preprint LMU-TPW
93-08, {\tt hep-th/9304034}. }
\lref\VafaQ{C. Vafa, ``Quantum Symmetries of String Vacua,''
{\it Mod. Phys. Lett.}
{\bf A4} (1989) 1615. }

\lref\HWPmirrors{D. Morrison, ``Picard-Fuchs Equations and Mirror Maps for
Hypersurfaces," In {\it Essays on Mirror Manifolds}, ed. S.--T. Yau,
(Int. Press Co., 1992) {\tt alg-geom/9202026}\semi
A. Font, ``Periods and Duality Symmetries in Calabi-Yau
Compactifications,'' {\it Nucl. Phys.} {\bf B391} (1993) 358, {\tt
hep-th/9203084}\semi
A. Klemm and S. Theisen, ``Considerations of One Modulus Calabi-Yau
Compactification: Picard-Fuchs Equation, K\"ahler Potentials and Mirror
Maps,"
{\it Nucl. Phys.} {\bf B389} (1993) 153, {\tt hep-th/9205041}.}
\lref\flops{P. Aspinwall, B. Greene and D. Morrison, ``Multiple Mirror
Manifolds and Topology Change in String Theory," {\it Phys. Lett.} {\bf 303B}
(1993) 249, {\tt hep-th/9301043}.}
\lref\flopsII{P. Aspinwall, B. Greene and D. Morrison, ``Calabi-Yau Moduli
Space, Mirror Manifolds and Spacetime Topology Change in String Theory,"
IAS and Cornell preprints IASSNS-HEP-93/38, CLNS-93/1236, to appear.}
\lref\cvetic{M. Cvetic, ``Exact Construction of (0,2) Calabi-Yau
Manifolds,'' {\it Phys. Rev. Lett.} {\bf 59} (1987) 2829.}
\lref\Miron{J. Distler, B. Greene, K. Kirklin and P. Miron, ``Calculating
Endomorphism Valued Cohomology: singlet spectrum in superstring models,"
{\it Comm. Math. Phys.} {\bf 122} (1989) 117.}
\lref\miracles{M. Dine and N. Seiberg, ``Are (0,2) Models String Miracles?,"
{\it Nucl. Phys.} {\bf B306} (1988) 137.}
\lref\GrPl{B.R. Greene and M.R. Plesser, ``Mirror Manifolds: A Brief
Review and Progress Report,'' {\tt hep-th/9110014}.}
\lref\DK{J. Distler and S. Kachru, ``(0,2) Landau-Ginzburg Theory,''
{\it Nucl. Phys.} {\bf B413} (1994) 213, {\tt hep-th/9309110}.}
\lref\masses{P. Candelas, X. De la Ossa and collaborators, to appear.}
\lref\eva{E. Silverstein and E. Witten, ``Global U(1) R-Symmetry and Conformal
Invariance of (0,2) Models," IASSNS-94/4,PUPT-1453, {\tt hep-th/9403054}.}
\lref\WittenElGen{E. Witten, ``Elliptic Genera and Quantum Field
Theory,'' {\it Comm. Math. Phys.} {\bf 109} (1987) 525\semi
``The Index of the Dirac Operator in Loop Space,'' in {\it Elliptic
Curves and Modular Forms in Algebraic Topology}, P.S. Landweber ed.,
Lecture Notes in Mathematics 1326 (Springer-Verlag, 1988).}

\lref\Mohri{T. Kawai and K. Mohri, ``Geometry of (0,2) Landau-Ginzburg
Orbifolds,'' KEK preprint, {\tt hep-th/9402148}.}
\lref\Nemeschansky{D. Nemeschansky and N. Warner, ``Refining the
Elliptic Genus,'' USC preprint, {\tt hep-th/9403047}.}

\newsec{Introduction}

Landau-Ginzburg orbifolds \refs{\Vafa, \Ken}\
describe special submanifolds in the moduli
spaces of Calabi-Yau models, at ``very small radius.''  New, stringy
features of the physics are therefore often apparent in the
Landau-Ginzburg
theories.
For example, these theories sometimes manifest enhanced gauge symmetries
which do not occur in the field theory limit, where the large radius
manifold description is valid.  And $\it all$ Landau-Ginzburg orbifolds
have one stringy symmetry, the quantum symmetry \VafaQ\ which
essentially counts the twisted sectors of origin of the various
elementary particles.
In spacetime, the quantum symmetry manifests itself as a discrete
R-symmetry.

Because the quantum symmetry is intimately related to the process of
constructing the orbifold, one expects that it will
correspond to an $\it exact$ symmetry in spacetime.
However, the spacetime theories arising from Heterotic string
compactification on \LG\ orbifolds typically correspond to $E_{6}\times
E_{8}$ gauge theories, and one also expects that
instanton effects associated with the observable $E_{6}$ and the
hidden $E_{8}$ might lead to a non-perturbative violation of various
discrete symmetries.
In fact, we shall see in \S2 that
the quantum symmetry of
many (2,2) (and more generally (0,2) \DK)
Landau-Ginzburg theories
seems naively to be broken by instantons of the observable gauge group.

This is somewhat puzzling: After all, the quantum symmetry is a
direct consequence of the orbifold construction.  If the string
description of our theory is correct, we really expect the symmetry to
be exact.  The resolution of this puzzle is provided in \S3, where we
recall that instanton solutions of string theory
must satisfy the Bianchi identity for the $H$ field.
In particular, this implies that in string theory,
instantons of the observable gauge group must
be accompanied by anti-instantons of the hidden gauge group,
gravitational instantons, or topologically non-trivial configurations of
the $H$ field.
This fact is quite important when
considering symmetry breaking effects of Yang-Mills instantons in string
compactifications.

In \S4, we show that the permitted configurations with an instanton of
the observable gauge group combined with an
anti-instanton of the hidden $E_{8}$ or a gravitational instanton
in fact preserve the quantum symmetry.
This involves a non-trivial conspiracy between the various gauge and
gravitational anomalies, and provides us with constraints on the
spectra of \LG\ orbifolds.
The solutions with
non-trivial $H$ field also preserve the quantum symmetry if one
assigns an appropriate transformation law to the model-independent
axion.
We discuss some implications of these results
in \S5.

Cancellation of ``duality anomalies'' has been
discussed in the context of
toroidal orbifolds by several groups \rdualanom.
The quantum symmetry of a Landau-Ginzburg theory is somewhat
analogous to the extra $\BZ_{2}$ symmetry\foot{Of course, this $\BZ_{2}$ is
just
a subgroup of the extra $SU(2)$ Kac-Moody symmetry which arises at
the self-dual radius.}
which arises at the fixed point of the
$R \rightarrow {1\over R}$
symmetry of the Teichmuller space of
circle compactifications \VafaQ.  In this sense and others,
our work represents a generalization of \rdualanom\ to the setting of
Landau-Ginzburg orbifolds.

In particular, Ib\'a\~nez and L\"ust (last reference of \rdualanom) used the
constraint of the cancellation of duality anomalies to restrict the possible
massless spectra which could arise in orbifold models. Using the constraints
we derive below, one could make similarly powerful statements about the
possible massless spectra of \LG\ orbifolds.

Indeed, certain ``bad models" (whose spectra of spacetime fermions already
looked peculiar) discussed in \DK\ violate the anomaly cancellation conditions,
and so can be ruled out on this basis.

\newsec{Quantum Symmetries and Anomalies}

The Landau-Ginzburg orbifolds are distinguished by the fact that they
all possess at least one discrete R-symmetry, namely the
quantum symmetry which counts the twisted sector $k$ of origin of
the various physical states.

For concreteness, let us focus attention on the quintic hypersurface in
$\cp{4}$. The \LG\ theory is a point of enhanced symmetry in the \Ka\ moduli
space. One normally says that the \LG\ orbifold has a $\BZ_5$
quantum symmetry, but
since one needs to include both NS and R sectors for the
left-movers, there are actually 10 sectors in the \LG\ orbifold \Us.
So one might
better think of the quantum symmetry as
$\BZ_{10}=\BZ_2\sdp\BZ_5$. Actually, this definition of the quantum symmetry
is a little awkward because the different components, under the decomposition
$E_6\supset SO(10)\times U(1)$, of a given $E_6$ representation transform
with different weight under this $\BZ_{10}$ symmetry. To fix this, we can
compose this symmetry with an element of the center of $E_6$, to obtain a
$\BZ_{30}=\BZ_3\sdp\BZ_{10}$ symmetry which acts homogeneously on $E_6$
multiplets. In the language of \Us, this $\BZ_{30}$ is generated by
\eqn\eSqdef{S_Q=\ex{2\pi i(3 k-2 \ql)/30}}
where $k=0,\dots,9$ labels the sector number of the \LG\ orbifold, and $\ql$ is
the left-moving $U(1)$ charge.\foot{The states found in
\Us\ were the massless spacetime fermions;  for a fermion which comes
from
the $(k+1)^{st}$ twisted sector, its scalar superpartner comes from the
$k^{th}$ twisted sector.}
The charges of the various massless
multiplets  under $S_Q$ are listed in Table 1 as integers $\in \BZ/30\BZ$.
In the table, $C$ and $R$ refer to the 101 complex structure moduli and
the \Ka\ modulus of the quintic.
$S$ and $S^{'}$ represent, respectively, groups of 200 and 24
other massless $E_{6}$ singlets, corresponding to elements of
$\rm {H^{1}}(\ET)$.
The former arise in the untwisted sector, the latter in a twisted sector of
the \LG\ orbifold \Us.

\def\tablerule{\omit&\multispan{14}{\tabskip=0pt\hrulefill}&\cr}
$$\vbox{\offinterlineskip\tabskip=0pt\halign{\hskip 1.0in
$#$\quad&\vrule #&\quad\hfil $\strut #$\hfil\quad &\vrule #&
\quad\hfil $#$\hfil\quad
&\vrule # &\quad\hfil $#$\hfil\quad &\vrule #&\quad\hfil $#$\hfil\quad
&\vrule #&\quad\hfil
$#$\hfil\quad &\vrule#&\quad\hfil $#$\hfil\quad &\vrule\ \vrule#
&\quad\hfil $#$\hfil\quad &\vrule #\cr
&\omit&{\bf 27}&\omit&\overline{\bf 27}
&\omit&C,S&\omit&R&\omit&S'&\omit&{\bf 78}&\omit&\CW&\omit\cr
\tablerule
S_Q&&-2&&8&&0&&6&&6&&0&&-6&\cr
\tablerule
\noalign{\bigskip}
\noalign{\narrower\noindent{\bf Table 1:} Charges ($\in \BZ/30\BZ$) of
the
spacetime matter multiplets,
gluons, and the spacetime superpotential $\CW$ under $S_Q$, the ``quantum"
R-symmetry present at the \LG\ point.
}
}}$$

On the world sheet, $S_Q$ simply enforces the fact that sector number is
conserved mod 10 in correlation functions. It is easy to see that, in {\it
spacetime},
$S_Q$ generates a discrete R-symmetry, under which the spacetime
superpotential has charge $-6$ mod 30. That is, one should add 3 to the
entries in Table 1 to obtain the $S_Q$-charge of the corresponding
right-handed fermions in these chiral multiplets, and {\it subtract} 3 to
obtain the $S_Q$ charge of the right-handed $E_6$ gluinos. Clearly
${\bf 27}^3$ and
$\overline{\bf 27}^3$ are couplings in the superpotential allowed by the
discrete
$R$ symmetry, whereas, say, ${\bf 27}^2\overline{\bf 27}^2$ is not.

This $R$-symmetry and its analogues in other (2,2) \LG\ theories are very
useful in proving the existence of a large class of (0,2) deformations.
The analogous quantum symmetries of the non-deformation (0,2) models
considered in \DK\ provide a stringent self-consistency check on the
assumption that those models are conformally invariant.  These matters
will be discussed elsewhere \DKtwo.

Our present interest is simply to check that this discrete $R$ symmetry is
nonanomalous.
It is easy to find the charge of
the $E_{6}$ instanton-induced 't Hooft
effective Lagrangian\foot{Though we have chosen to discuss the anomaly in
terms of the (nonperturbative) 't Hooft effective Lagrangian, one could
equally well phrase the discussion of the anomaly in terms of the noninvariance
of the (nonlocal) one loop effective Lagrangian. That is, if the anomaly
doesn't vanish, the symmetry is broken already in string {\it perturbation}
theory.}\ ${\cal L}_{eff}$ \Gt\ under $S_{Q}$:

\eqn\thcharge{ {\rm {Charge~ of}}~ {\cal L}_{eff} ~=~
2~\cdot~\left[(-3)\cdot c({\bf 78}) + 101\cdot(1)\cdot c({\bf 27}) + (11)\cdot
c(\overline{\bf 27})~\right]~{\rm mod}~30}
$$ = 600 ~{\rm mod}~30~= 0 \ {\rm mod}\ 30 $$
where $c({\bf 78})=12$, $c({\bf 27})=c(\overline{\bf 27})=3$ for $E_6$.

So we see that, for the quintic, the quantum $R$-symmetry is
not broken by $E_{6}$ instantons \Us.
Unfortunately, this will be far from true in many other examples.

In the case of more general (2,2) or (0,2) theories, the quantum
$R$-symmetry is generated by
\eqn\eSQgen{S_Q=\ex{2\pi i(kr -2\ql)/2mr}}
where $k=0,1,\dots 2m-1$ is the sector number, and $r$ is the ``rank of
the
vacuum gauge bundle" -- $r=3,4,5$ for spacetime
gauge group $G=E_6,~SO(10),$ or $SU(5)$.  In particular, $r=3$ for all
(2,2) theories.
We are interested in checking the transformation
properties of the 't Hooft effective
Lagrangian induced by $G$ instantons under \eSQgen, just as we did
for the quintic.  If ${\cal L}_{eff}$ is not invariant, then
instantons break the quantum symmetry --
that is, it is anomalous.

Rather generally, the computation of
the transformation properties of the 't Hooft effective Lagrangian can
be rephrased as the computation of a
new {\it index} for \LG\
orbifolds. As usual, the index can be phrased as a trace over massless
($L_0=\bar L_0=0$) states in the right-moving Ramond sector
\eqn\newanom{{\rm A_{1}}=
\sum_{k}Tr_{R}\left({\ql^{2}\over {2r}}\Big({rk}- 2\ql\Big)(-1)^{
F_R}\right)\quad {\rm mod}\ 2mr}
\eqn\newanomtwo{= \sum_{k} Tr_{R}\Big( {\ql^{2}\over 2}k(-1)^{F_{R}} \Big)\quad
{\rm
mod}\ 2mr}
where, again, $k$ is the sector number, $\ql$
is the left-moving U(1) charge, and
$F_R=\qr+1/2$ is the right-moving fermion number.
The factor of ${1\over 2r}$ in \newanom\ is necessary because the
worldsheet $U(1)$ current $J$ has been normalized so that
$J(z) \cdot J(w) \sim {r\over {(z-w)^{2}}}$, while the current
which goes into the vertex operator
for the gauge boson is canonically normalized to have a central
term of ${1/2\over {(z-w)^{2}}}$.
One easily finds \newanomtwo\ from \newanom\ by realizing that
$\sum_{k} Tr_{R} \Big(\ql^{3}(-1)^{F_{R}}\Big)$
measures the gauge anomaly
which vanishes.
Note that one should include the $(16-2r)$ free left-moving Majorana-Weyl
``gauge'' fermions when taking
the sum \newanomtwo.
For example in an $E_{6}$ theory, the various $E_{6}$
representations
can be decomposed as representations of the maximal $SO(10) \times U(1)$
subgroup, and each state of the \LG\ theory should be multiplied by the
dimension of the $SO(10)$ representation of the corresponding spacetime
particle.
If the index \newanomtwo\ doesn't
vanish, then the quantum R-symmetry is broken by instantons.

The index that we have introduced is but the first term in a modular
form
\eqn\elliptic{F(q)=\sum_{k}Tr_{R}\Big({{\ql^{2}\over 2}}
k (-1)^{
F_R}q^{L_0-c/24}\Big)}
with coefficients in $\BZ/2mr\BZ$.
This is a refinement for \LG\ orbifolds of the elliptic genus
\refs{\WittenElGen,\WitMin,\Mohri,\Nemeschansky}.

We find that in many models, including well known (2,2) compactifications,
the quantum symmetry $\it is$ broken by instantons of the observable
gauge group.
Consider, for example, the (2,2) Landau-Ginzburg theory with
superpotential
\eqn\anomsup{W = \sum_{i} (S_{i}^{3} + S_{i}T_{i}^{3}) + S_{4}^{3}~.}
This is the Landau-Ginzburg description of the $(1,16,16,16)$ Gepner
model \Gepner\ with exceptional modular invariants for the $16$s.
The Calabi-Yau phase of this theory was first discussed by Schimmrigk
\Schimmrigk.
This model has been of some interest in the literature
because, after suitable orbifolding, it yields a quasi-realistic three
generation model.  The quantum symmetry of the
string theory compactified on \anomsup\
is, naively,  broken by $E_{6}$ instantons.

In the language we have been using, the quantum symmetry of \anomsup\
is generated by
\eqn\anomsq{S_Q=\ex{2\pi i(3 k-2 \ql)/54}~.}
To see that this symmetry is anomalous, recall that the compactification
on \anomsup\ yields 35 left-handed generations (${\bf 27}$s of $E_{6}$) and 8
left-handed anti-generations ($\overline{\bf 27}$s of $E_{6}$).  In the
orbifold, the 35 left-handed generations have $3k-2\ql = -1$.  Six of the
left-handed anti-generations have $3k-2\ql = 19$, while one has $3k-2\ql =
13$ and one has $3k-2\ql = 25$.  Remembering also that the left-handed
gaugino in the ${\bf 78}$ of $E_{6}$ has $3k-2\ql = 3$, and using
$c({\bf 27})=c({\overline{\bf 27}})=3,~c({\bf 78})=12$, we see that the
total anomaly
is\foot{Actually, by doing the sum over only left-handed fermions, we
are computing half of $A_{1}$ and requiring that it vanish $\rm
{mod}~mr$.}
$$(3)\cdot 12+35 \cdot(-1)\cdot 3 + \left(6\cdot(19) +
(13) +  (25)\right)\cdot 3 = 387 = 9\ {\rm mod}~27~.$$
Therefore, the quantum
R-symmetry is broken by instanton effects in the spacetime theory.

There are many simpler examples which exhibit this anomaly.  For example
it is a simple matter to compute
\newanom\ for the $r=3$ theories corresponding to the (2,2)
compactifications on the Calabi-Yau hypersurfaces in
$W\BP^{4}_{1,1,1,1,3}$ and
$W\BP^{4}_{1,1,1,1,4}$.  In the first case, with $m=7$, one finds that
$A_{1}$ is equal to $324~{\rm mod}~42 \neq 0$ and that the quantum
$\BZ_{7}$ symmetry is completely broken by instantons.  A similar
computation shows that in the second case
the quantum $\BZ_{8}$ is
broken to a $\BZ_{2}$.

The fact that non-perturbative effects seem to be breaking the quantum
symmetry of these \LG\ orbifolds is very disturbing -- one would expect,
because of its role in constructing the orbifold, that the quantum
symmetry should be an $\it exact$ symmetry of the compactified string
theory.  We shall see in \S4 that in fact it is -- but to understand
why
string theory evades the effects of $E_{6}$ instantons, we must first
remind ourselves of some pertinent facts about
anomaly cancellation.

\newsec{The Green-Schwarz Mechanism and Stringy Instantons}

The cancellation of spacetime and worldsheet anomalies is one of the
most delicate features of string theory.  The stringy modification
of the minimal $N=1, D=10$ supergravity theory coupled to super
Yang-Mills theory which allows for the cancellation of all gauge and
gravitational anomalies \GS\ involves assigning non-trivial gauge and
local Lorentz transformation properties to the antisymmetric tensor
field $B$.  Under an infinitesimal gauge transformation with parameter
$\Lambda$ and an infinitesimal local Lorentz transformation with
parameter $\theta$, the $B$ field transforms as
\eqn\Btrans{\delta B = tr (Ad\Lambda) - tr (\omega d\theta)}
where $A$ is the gauge connection and $\omega$ is the spin
connection.
The transformation law \Btrans\ which cancels the space-time anomalies
is also responsible for the cancellation of $\sigma$-model anomalies
in the heterotic string, which may occur when one considers nontrivial
maps of the worldsheet to spacetime \WitGlob.

\Btrans\ implies that the gauge invariant field strength $H$ of $B$
obeys the Bianchi identity
\eqn\modH{dH = tr F_1^{2}+tr F_2^{2} - tr R^{2}~.}
Here
$H$ is a 3-form and $F_{1,2}$ and $R$ are (Lie-algebra valued) two-forms in
spacetime.  $tr F_i^{2}$ and $tr R^{2}$ represent the first
Pontryagin classes $p_{1}(V_i,\BR)$ and $p_{1}(T,\BR)$ of the
$E_{8}\times E_{8}$ gauge bundle $V_1\oplus V_2$ and the tangent bundle $T$
of our 4-manifold.   This equation leads to an important topological
restriction for string propagation on
compact spaces, which is closely related to cancellation of
local and global anomalies in string theory \refs{\WitGlob, \WitSome}.

The case of interest to us is strings propagating in
four-dimensional Minkowski space.  In order to satisfy \modH, we can
consider at least three types of configurations.

Let us first try to satisfy \modH\ using only gauge instantons.
Actually, when searching for such instanton solutions, we
should consider a Euclidean continuation and
compactification of Minkowski space to a four-dimensional torus.
The relevant property of this space for us is that $tr\, R^{2} = 0$.
So
the condition \modH\ reduces to
\eqn\reduced{tr\, F_{1}\wedge F_{1} + tr\, F_{2}\wedge F_{2} = 0}

Of course a single instanton of either the first or second $E_{8}$
cannot possibly satisfy \reduced!  So we see that the instanton
configurations of the observable $E_{6}$, whose effects we previously
considered, will never appear by themselves in string theory.  They must
be accompanied by anti-instantons of the hidden $E_{8}$, in such a
way that \reduced\ is satisfied.

We can also find solutions
of \modH\ which combine gauge
instantons of the observable gauge group with
gravitational
instantons in such a manner that
\eqn\grav{tr\, R\wedge R = tr\, F_{1}\wedge F_{1}}
is satisfied.
Similarly, solutions of this sort with the gauge instantons of the
observable gauge group replaced by those of the hidden $E_{8}$ must be
considered.

Finally, we can build solutions of \modH\ in which we take a single
instanton of the observable gauge group, and choose a non-trivial
configuration of the $H$ field in such a way that \modH\ is still
satisfied.  This is only possible in noncompact (Euclidean) spacetimes.
Such configurations have been discussed in \Rohm\ in $\BR^4$
and more recently have been reincarnated as ``gauge fivebranes''
\Callan, which is what we will call them.  In fact, the
configurations which satisfy \reduced\ can presumably be constructed
by taking two widely separated gauge fivebranes, one built around an
instanton of the observable gauge group and the other built around an
anti-instanton of the hidden gauge group.

One could also study configurations with non-trivial $H$
fields, gauge, and gravitational instantons \Callan;
we shall only consider the
more basic types of solutions discussed above.  If those
basic configurations do not break a symmetry, the more elaborate
solutions will also preserve it.

Therefore, the
questions of interest become:  What are the symmetry
breaking effects of
instanton-anti-instanton pairs, with an instanton from the observable
$E_{6}$ and an anti-instanton from the hidden $E_{8},$
what are the effects of gauge
instanton/gravitational instanton pairs, and what are the effects of
gauge fivebranes?

\newsec{Quantum Symmetries Revisited}

The computation of quantum symmetry breaking due to instantons
of the observable
gauge group has already been discussed in \S2.  The effect of
hidden $E_{8}$
instantons on the quantum symmetry is even easier to analyze.

The only fermi fields charged under the hidden $E_{8}$ which arise in
twisted sectors of the \LG\ orbifold are the gluinos in the ${\bf 248}$ of
$E_{8}$; the left-handed gluinos come from the $k=1$ twisted sector \Us.
Therefore, the 't Hooft effective Lagrangian induced by instantons of
the second $E_{8}$ has $S_{Q}$ charge
\eqn\secondcharge{A_{2}=-2 \cdot c({\bf 248}) \cdot r ~{\rm {mod}~}2mr ~=~
-60r~{\rm
{mod}}~2mr~.}

Considering only configurations with $E_{8}$ anti-instantons for every
instanton of the observable gauge group, we see that the resulting
effective Lagrangian will then have $S_{Q}$ charge
\eqn\anomsum{A_1-A_2 = 60r + \sum_{k} Tr_{R}\Big(
{\ql^{2}\over 2}k(-1)^{F_{R}} \Big) \quad {\rm mod}\ 2mr~.}
It is easy to check that \anomsum\ in fact vanishes for the theories
whose quantum symmetries appeared to be broken by
instantons of the observable
gauge group.
For example, for the 27 generation $(1,16,16,16)$ Gepner model discussed
in \S2, one sees that \anomsum\ is equal to
$-774 + 180 ~{\rm mod}~ 54 = -594~{\rm mod} ~54 = 0$.
Similarly, for the hypersurface in $W\BP^{4}_{1,1,1,1,3}$, \anomsum\
is equal to $324 + 180 ~{\rm mod} ~42 = 504 ~{\rm mod}~ 42 = 0.$
In fact, the anomaly cancellation works in all of the theories we have
checked.

Similarly, one finds that a single gravitational instanton contributes
(up to an overall normalization)
\eqn\gravinst{A_{3}=
-452r + \sum_{k} ~Tr_{R}\left((rk-2\ql)(-1)^{F_{R}}\right)~
{\rm mod}~
2mr}
to the anomaly.  The $-452r$ comes from the contribution of the $E_{8}$
gauginos ($-496r$), the dilatino ($2r$), and the gravitino ($42r$) (to
fix the contributions it is enough to know that the left-handed gauginos
and gravitino and the right-handed dilatino come from the $k=1$ twisted
sector).
The
fact that the gravitino has the same chiral anomaly as $-21$ Weyl
fermions is familiar from the study of gravitational index theorems
\Duff.  As in \newanom, one must include the $16-2r$ free ``gauge''
fermions when taking the sum \gravinst; their sole effect is to multiply
the contribution of a fermion transforming in the ${\cal R}$
representation of the observable gauge group by ${\rm dim}({\cal R})$.

So to summarize, the
full anomaly in the quantum symmetry is proportional to
\eqn\fullanom{\left(- {\textstyle{1\over 24}}A_{3} p_1(T)+
A_{1}p_1(V_1)+A_{2}p_1(V_2)\right)~ {\rm mod} ~ 2mr.}
The normalization of the gravitational contribution to the anomaly
relative to the gauge contribution is the same $-{1\over 24}$
which appears in the
Atiyah-Singer index theorem for the twisted spin complex \Hanson.

We are interested in the vanishing of \fullanom\ for instanton
configurations with gravitational instanton number $n_{1}$,
$n_{2}$ observable gauge group instantons, and $n_{3}$ hidden $E_{8}$
instantons.
The configurations which satisfy \modH\ with trivial $H$ field
necessarily have
\eqn\constr{p_1(T) = p_1(V_1)+p_1(V_2)~.}
However, compact spin four-manifolds
without boundary
have
\eqn\gravinst{p_1(T) \in 24\BZ~.}
That is, the index of the Dirac operator on such a manifold is an integer.

Therefore, to insure that the solutions with non-trivial gauge and
gravitational configurations do not violate the quantum symmetry, one
needs to check
\eqna\anomtwo
$$\eqalignno{                      A_1-A_2&=0~ {\rm mod}~ 2mr&\anomtwo a\cr
                        A_{3} - 24 A_{2} &= 0~ {\rm mod}~ 2mr&\anomtwo b\cr
                        A_{3} - 24 A_{1} &= 0~ {\rm mod} ~2mr~.&\anomtwo c\cr}
$$
These are simply the conditions for the
cancellation of the anomaly in the case
where one of the three first Pontryagin classes in \constr\
vanishes. If $\anomtwo{a,b,c}$ are satisfied, then
the quantum symmetry is non-anomalous for {\it any} choice of instanton
satisfying
\constr.\foot{In fact, the
conditions $\anomtwo{a,b,c}$ are not all independent;
one can see that if $\anomtwo{a}$ and either of
$\anomtwo{b,c}$ are satisfied, then
all three conditions are satisfied.}

We have checked that $\anomtwo{a,b,c}$ are
satisfied
by all the models that we have
studied.
For example, in the \LG\
phase of the hypersurface in $W\BP^{4}_{1,1,1,1,3}$,
one finds $A_{3} \simeq 6~{\rm mod}~42$ while
$A_{1} \simeq 30~{\rm mod}~42$
and $A_{2} \simeq 30~{\rm mod}~42$, so $\anomtwo{a,b,c}$ are indeed
satisfied.

The effects of gauge fivebranes, or
solutions with non-trivial $H$ field
in general, are
slightly more subtle.  Recall that
the 't Hooft effective Lagrangian generated by an instanton in string
theory has the general form
\eqn\leff{{\cal L}_{eff} \sim \ex{-8\pi^{2}/g^{2}}\ex{ia/f_{a}}~
\psi_{1}\dots\psi_{N}}
where $a$ is the model-independent axion, $g$ is the string-coupling,
and the $\psi_{i}$ denote some space-time fermion fields.
So notice that even if ${\cal L}_{eff}$ seems naively (based on the
transformation properties of the $\psi_{i}$) to violate the quantum
symmetry, by requiring that the axion
$a$ transform by an appropriate shift one can always restore invariance
of ${\cal L}_{eff}$.\foot{This method has been
employed by other authors in
similar contexts \rdualanom \BanksD.}
Note that in order for this trick to work, it is crucial that the full anomaly
be proportional to $dH=tr F_1^2 + tr F_2^2 - tr R^2$. That is, it is crucial
that the anomaly \fullanom\ vanish whenever \constr\ is satisfied,
or equivalently that  $\anomtwo{a,b,c}$ all vanish.
Only then
can one assign a transformation law to $a$, which cancels the anomaly.

But assuming that the anomalies vanish, as they do in the sensible (2,2)
and (0,2) theories we have checked, one can also cancel the symmetry
breaking effects of gauge fivebranes by requiring that $a$ transform by
a shift under the quantum symmetry.
This has an interesting consequence for the dilaton-axion
moduli space, which we mention in \S5.

\newsec {Discussion}

We have seen that
that one encounters certain puzzling facts
when one naively extrapolates field theory non-perturbative
effects to a stringy context.  These
puzzles are naturally resolved when proper
account is taken of the condition \modH, which solutions of
string theory must satisfy.

The vanishing of the anomalies $\anomtwo{a,b,c}$
for Landau-Ginzburg
orbifolds is non-trivial.  It is giving
us some interesting constraints on the twisted sectors the matter fields
and moduli must come from in \LG\ compactifications.  For instance, in
the quintic, it is
crucial
in the anomaly cancellation
that 24 of the $E_{6}$ singlets related to neither complex
structure nor \Ka\ structure deformations arise from the $k=3$ twisted
sector
instead of the $k=1$ sector, where one might have naively expected
to find them \Us.  Furthermore, in models like the 27-generation
model and the $W\BP^4_{1,1,1,1,3}$ model, where ``extra" singlets arise at the
\LG\ radius, both the number of such extra singlets and the twisted sectors
in which they appear are constrained by the cancellation of the gravitational
anomalies $\anomtwo{b,c}$. Indeed, were these extra singlets {\it not} to
arise, then the anomalies would not cancel in these models!

Clearly, this situation is much more general than \LG\ theories, or the
orbifold theories studied in \rdualanom. Rather
generally, one has a moduli space $\CM$ of string vacua, which is obtained
as the quotient of the ``naive" moduli space by the action of a modular
group $\CG$. In general, $\CM$ contains orbifold ``points," which are
left fixed by various subgroups $\CG_0\subset\CG$. $\CG_0$ then acts as an
automorphism of the  conformal field theory corresponding to the fixed point
set. In favourable circumstances, one even generates all of $\CG$ in this
fashion. Demanding that these symmetries be nonanomalous, {\it i.e.}~that the
modular group $\CG$ be a good symmetry of string theory, is a powerful
restriction on the spectrum of the string theory. It constrains not just the
states which are massless generically in $\CM$, but also those states which
become massless only at the  fixed points.

The fact that, in those models where $A_{1,2,3}$ do not separately vanish mod
$2mr$, one must transform the model-independent axion
under the quantum symmetry is also interesting.
Roughly speaking,
it means that the T-duality which acts on the \Ka\ moduli
space also acts non-trivially on the dilaton.  So the true moduli space
is a quotient
$$\CM={\CD\times\CR\over\BZ_m}$$
of the product of the dilaton and \Ka\ moduli spaces, where the quantum
symmetry acts nontrivially on both factors.
For believers in
strong-weak coupling duality (see
\Schwarz\ and \Asen\
and references therein), this means that beyond the $SL(2,\BZ)$
symmetry which acts on the dilaton-axion modulus alone, one must further
quotient the dilaton-axion moduli space
by a $\BZ_{m}$ action. Since, in general, these do not commute\foot{The
$SL(2,\BZ)$ symmetry is generated by $\tau\to \tau +1$ and $\tau\to
-1/\tau$, where $\tau={a\over2\pi f_a} + {4\pi i\over g^2}$. The $\BZ_m$
symmetry acts on $\tau$ by $\tau\to\tau+ n/m$, for some $n$.}, the full modular
group is a semidirect product.

We also noted before that, if the anomalies
 $\anomtwo{a,b,c}$ do not vanish for a given
compactification, then the quantum symmetry will be broken (one cannot
cancel \fullanom\ by shifting the model-independent axion).  Although
these anomalies do indeed vanish for all the
(2,2) theories that we have checked, they do not vanish
automatically for (0,2) theories.
The anomaly cancellation conditions are satisfied, in the
cases we have checked, for the (0,2) theories with ``generically
sensible'' spectra, like the $Y_{W5;4,4}$ model
discussed in \DK.

However, for models whose spectra
exhibit unexpected pathologies like the $Y_{W4;10}$ model in \DK,
\anomsum\ does not always vanish.  In particular, the $Y_{W4;10}$ model
suffers from an anomalous quantum symmetry.  We are tempted to conclude that
this is yet another way of seeing that this ``bad model" does not give rise
to a consistent string vacuum. This is a {\it stronger} conclusion than
was ventured in \DK. There, we noted that for very special choices of the
defining data, we could obtain a sensible-looking spectrum
for a \LG\ theory with
unbroken $E_6$ gauge symmetry. But now we see that even that theory suffers
from an anomaly in the quantum symmetry. Hence,  since
the quantum symmetry should be a good symmetry of string theory, we must
reject the $Y_{W4;10}$ model, even at its $E_6$-symmetric point.

\bigbreak\bigskip\bigskip\centerline{{\bf Acknowledgments}}\nobreak
\frenchspacing{
We would like to thank E. Silverstein, C. Vafa and E. Witten for discussions.
This work was supported by NSF grant PHY90-21984
and by the A.P. Sloan Foundation. }

\listrefs
\end